# Three-dimensional chiral microstructures fabricated by structured optical vortices in isotropic material

**Jincheng Ni, Chaowei Wang, Chenchu Zhang, Yanlei Hu\*, Liang Yang, Zhaoxin Lao, Bing Xu, Jiawen Li, Dong Wu\* & Jiaru Chu**

CAS Key Laboratory of Mechanical Behavior and Design of Materials, Department of Precision Machinery and Precision Instrumentation, University of Science and Technology of China, Hefei, Anhui 230027, China

**\***Email: huyl@ustc.edu.cn and dongwu@ustc.edu.cn

## Abstract

Optical vortices, as a kind of structured beam with helical phase wavefronts and 'doughnut' shape intensity distribution, have been used for fabricating chiral structures in metal and spiral patterns in anisotropic polarization-dependent azobenzene polymer. However, in isotropic polymer, the fabricated microstructures are typically confined to non-chiral cylindrical geometry due to two-dimensional 'doughnut' intensity profile of optical vortices. Here we develop a powerful strategy for realizing chiral microstructures in isotropic material by coaxial interference of a vortex beam and a plane wave, which produces three-dimensional (3D) spiral optical fields. This coaxial interference beams are creatively produced by designing the contrivable holograms consisting of azimuthal phase and equiphase loaded on liquid-crystal spatial light modulator. Then, in isotropic polymer, 3D chiral microstructures are achieved under illumination of the coaxial interference femtosecond laser beams with their chirality controlled by the topological charge. Our further investigation reveals that the spiral lobes and chirality are caused by the interfering patterns and helical phase wavefronts, respectively. This technique is simple, stable, and easy-operation, and offers broad applications in optical tweezers, optical communications and fast metamaterial fabrication.

Optical vortices as a kind of structured light beams[1] have led to wide-ranging applications in micromanipulation[2-4], free-space communications[5,6] and micro/nanofabrication[7,8] over the past couple of decades. In 1992, it was Allen[9] recognized that an optical vortex with an $e^{il\varphi}$ azimuthal phase, where $l$ is an integer, called topological charge, has helical phase wavefronts and carries an orbital angular momentum (OAM) of $l\hbar$ ($\hbar$ is the reduced Planck's constant). In the next decade, optical vortices with dark focuses are used as unconventional optical tweezers in trapping objects without damage and repellence[3]. The trapped colloidal particle translates around the 'doughnut' shape intensity circumference, which is a visual way to observe OAM transferred from optical vortices to drive the particles[2].

As demonstrated in recent studies, optical vortices have also been contributed to achieve complex chiral structures. For example, Toyoda et.al.[8,10] demonstrated that circularly polarized optical vortex can twist metal to form chiral nanostructures by transferring light helicity to nanostructures, where OAM forces the melted metal to revolve around the axis of optical vortex and optical scattering force directs it to the core. Meanwhile, Ambrosio et.al.[7] presented that spiral-shaped relief patterns are manufactured on an anisotropic polarization-dependent azobenzene-containing polymer film under the illumination of tightly focused vortex beams with linearly polarization. This phenomenon ascribed to slight mass-transport on azimuthal direction arises from the surface-mediated interference of longitudinal and transverse components of optical field. However, in isotropic polymers, non-chiral cylindrical microstructures corresponding to 2D 'doughnut' shape intensity distribution are typically fabricated under illumination of optical vortices[11] or high-order Bessel beams[12], which carries OAM. Therefore, it is of great interest to develop a novel and simple method to fabricate 3D chiral microstructures in isotropic material by optical modulation of vortex beams.

Femtosecond laser two-photon polymerization (TPP) based on direct laser writing is regarded as one of the most promising methods for fabricating flexible 3D micro/nanostructures, such as photonic crystals[13], microchips[14], metamaterials[11,15] with high resolution (< 100 nm). However, a complex 3D structure is time-consuming due to the point-to-point writing strategy with a tightly focused laser[16]. A 3D chiral structure polymerized in a single pulse without laborious scanning plays a crucial role in fast fabricating microstructures[17].

In this work, we propose a simple and effective approach to produce 3D controllable chiral microstructures inside an isotropic polymer with the interference beam of helical phase wavefronts and plane waves. First, we creatively design interfering vortex holograms (IVHs), which could generate interfering vortex beams (IVBs) of optical vortices and plane waves through liquid-crystal spatial light modulator (SLM). Then we demonstrate that IVBs with varying topological charge are used for generating 3D chiral microstructures with $l$ spiral lobes in an isotropic material. The cross-sectional spiral lobes are corresponding to the interference patterns, and the chirality are induced by the helical phase wavefronts. Our technique, without the time-consuming scanning process, shows distinct capability for the fabrication of 3D large-area (> 5 mm$^2$) chiral microstructures with ~100 nm precision.

**Results**

**Holographic generation of 3D spiral optical field by coaxial interference of a vortex beam and a plane wave.** Optical vortex can be conveniently created after a plane wave reflecting from a SLM, which is loaded an azimuthal phase hologram[7,18]. Computer generated hologram (CGH) with an azimuthal phase $mod(l\varphi, 2\pi)$ is corresponding to the helical phase wavefronts and a 2D 'doughnut' shape intensity distribution, as illustrated in Fig. 1a. $\varphi$ is the azimuthal angle around the optical axis and topological charge $l$ could be any positive or negative integer. With a given $l$, vortex beam has $l$ intertwined helical phase fronts, which has left-handedness (or right-handedness) with positive (or negative) $l$, and a particular annular intensity profile, whose radius increases with $l$. Specially, when $l$ equals zero, SLM with displaying equiphase distribution CGH acts as a mirror, and the reflected laser beam would be a plane wave[18]. As we know, the topological charge $l$, or the state of OAM, of optical vortices can be visually revealed by interfering with a reference beam[19,20]. Figure 1b shows an optical vortex interfered with a plane wave generates spiral intensity interferogram on the focal plane. There are $l$ spiral lobes in the interference patterns with topological charge $l$, while the sign of topological charge $l$ determinates the handedness of the spiral lobes. However, the conventional interference set-up, where a Gaussian reference beam is split and then superimposed with the OAM beam, generally requires a steady mechanics because the interference pattern is sensitive to the phase of Gaussian reference beam[20,21]. In order to avoid the variation of relative phase between optical vortices and Gaussian reference beam, we adopt SLM to generate the stable spiral patterns through copropagating interference of optical vortices and plane waves for the first time.

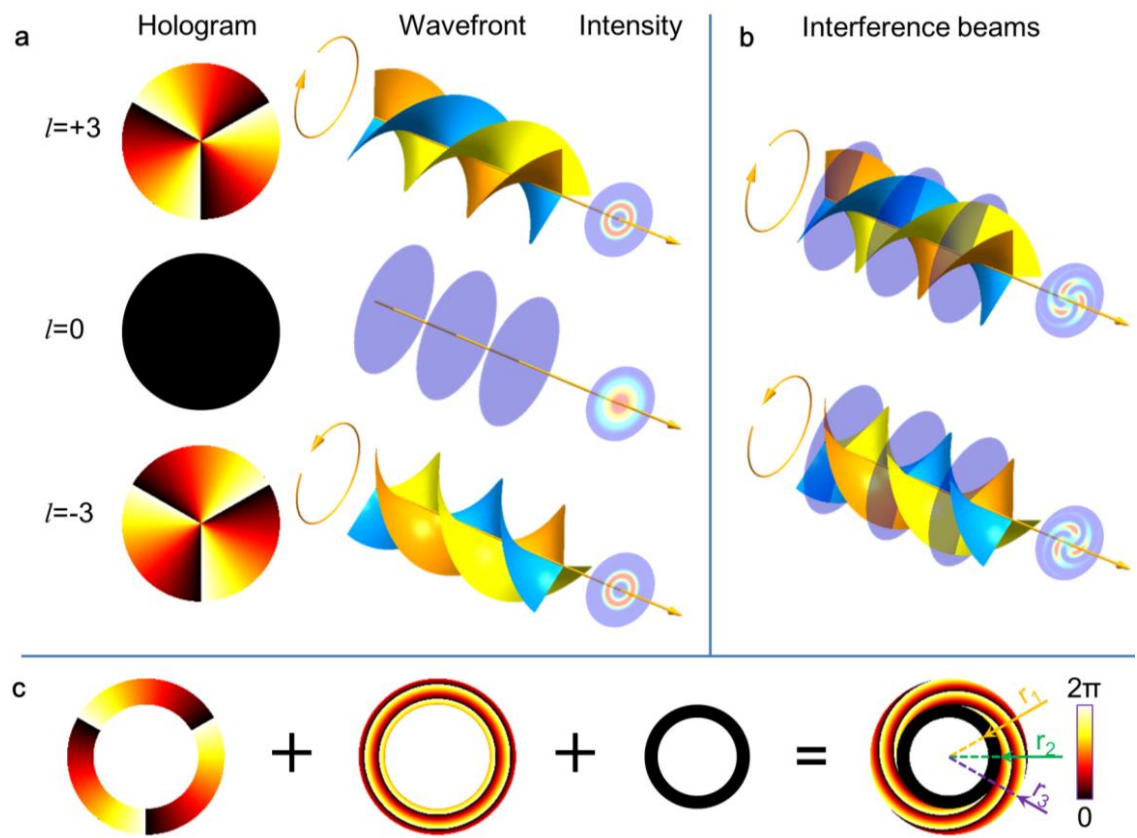

**Figure 1 | Holographic generation of 3D spiral optical field by coaxial interference of a vortex beam and a plane wave.** (**a**) Illustration of the optical vortices holograms with topological charge $l = 3$ (top panels),-3 (bottom panels) and its corresponding optical helical phase wavefronts and 'doughnut' shape intensity distributions. A regular Gaussian beam hologram (middle panels) with $l = 0$ has a plane wavefront. (**b**) Interfering a vortex beam ($l = 3$ or -3) with a plane wave results in a spiral pattern. (**c**) Azimuthal phase hologram ($l = 3$) is superimposed with annular zero equiphase in the central area. Radial shift phase tilts the optical vortex to interfere with the plane

wave.

Figure 1c explains the procedure of designing IVH with $l = 3$, consisting of both azimuthal phase and equiphase for 3D spiral optical fields. Annular hologram of an optical vortex superimposes zero phase value in the central area, which is regarded as a plane wave, and radial shift on vortex hologram realizes interfering with the plane wave[22]. The IVHs can be presented as

$$\emptyset(r,\varphi) = \begin{cases} 0 & r_1 < r \leq r_2, 0 < \varphi \leq 2\pi; \\ (l\varphi + 2\pi\frac{r}{R}) \bmod 2\pi & r_2 < r \leq r_3, 0 < \varphi \leq 2\pi; \\ 2\pi & else. \end{cases} \quad (1)$$

where $\emptyset(r,\varphi)$ is the transverse phase distribution on IVHs, $r$ is the radial distance from the hologram center and $R$ is radial shift value, which determines the shift angle of vortex beam. Note that, the size of central zero area is tailored accurately to get a perfect intensity pattern (see Figs. S1, S3).

**Experimental and theoretical intensity distributions of annular optical vortex (AOV) and IVBs with different spiral lobes.** Annular optical vortex (AOV) superimposed with a radial shift phase can shrink the 'doughnut' shape intensity profiles for interfering the plane wave (see Fig. S2). The IVBs generated by IVHs feature the interference patterns of optical vortices and plane wave references (Figs. 2a-l). Figure 2m shows the simulated and detected optical intensity distributions of IVBs with varying topological charges $l$ = ±3, ±4 and ±6. The observed intensity profiles of IVBs are caught by a CCD on plane A, where the distance from L$_3$ is $d = 350$ mm (see Fig. 3a). These measurements are in good agreement with the corresponding simulations. The topological charges of the IVBs are obvious to distinguish without another beam for interference. Furthermore, radii of the interference patterns with different topological charges are almost the same except the dark gaps on the 'doughnut' shape profile (Figs. 2f, i). The dark gaps with zero intensity are corresponding to the zero phase area of azimuthal phase in IVHs, and the dark gaps number represents the magnitude of topological charge and the twist direction indicates its sign.

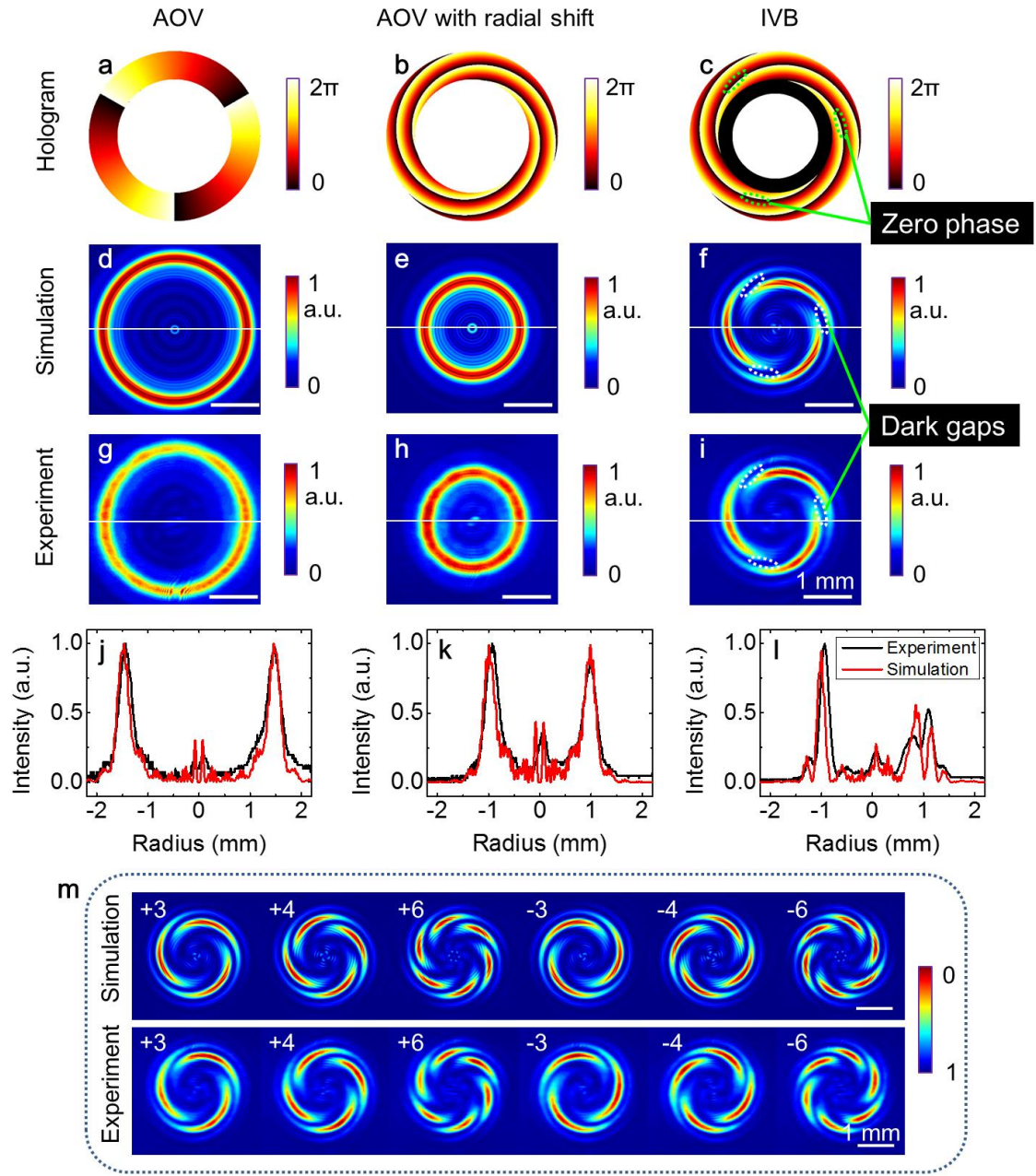

**Figure 2 | Experimental and theoretical intensity distributions of annular optical vortex (AOV) and IVBs with different spiral lobes**. Annular azimuthal hologram (**a**) has the simulated (**d**) and experimental (**g**) 'doughnut' shape intensity distributions. Annular azimuthal hologram with a radial shift (**b**) has a constringent 'doughnut' shape intensity profile as seen from the simulation (**e**) and experiment (**h**). IVH (**c**) has the simulated (**f**) and experimental (**i**) spiral shape intensity profiles. (**j-l**) The comparison of vertical cuts of intensity distributions between simulation and experiment corresponding to (**a-c**). (**m**) Simulated intensity profiles and observed results of IVBs with topological charge $l = \pm 3$, $\pm 4$ and $\pm 6$. The dark gaps in simulated (**f**) and experimental (**i**) intensity patterns of IVBs are corresponding to the zero phase in hologram (**c**).

**SLM-based experimental set-up for 3D chiral microstructures in isotropic material.** The experimental set-up used for fabricating chiral microstructures is depicted in Fig. 3a. We use a femtosecond laser working at wavelength $\lambda = 800$ nm, which is adjusted to be a linearly polarized Gaussian beam. The beam reflected from SLM displaying IVH, has helical phase wavefronts and spiral lobe patterns. In order to remove spiral patterns from other diffraction orders and control the desirable spatial shape of the reflecting beam, an annular tilted shift phase hologram is added to the holograms (see Fig. S4)[23]. The sample is inverted under a 100× objective lens (NA 0.9) of the microscope system (see Methods). The exposure time is 4 s and the laser power measured after the iris is 100 mW. A 3D chiral microstructure is polymerized in the photoresist on the defocused area, illustrated in Fig. 3b.

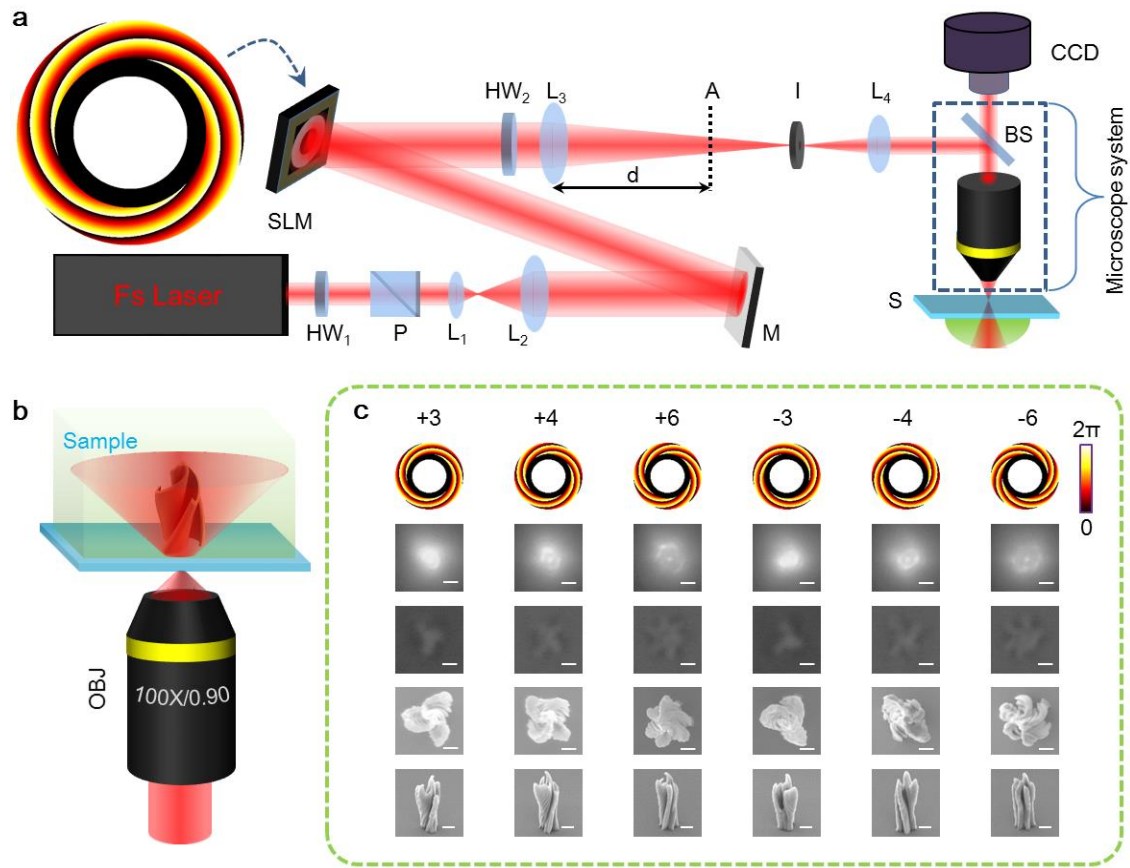

**Figure 3 | SLM-based experimental set-up for 3D chiral microstructures in isotropic material.** (**a**) The femtosecond laser is expanded by a telescope (lenses $L_1$-$L_2$) to match the aperture of SLM, after its polarization and power controlled with a half wave plate ($HW_1$) and a polarizer (P). The Gaussian beam, reflected by a mirror (M), is modulated by a SLM, which displays a designed hologram. Subsequently, the desired first-order diffraction beam is selectively introduced into the

microscope system, via an iris (I) positioned at the focal plane of lenses ($L_3$ and $L_4$). A half-wave plate ($HW_2$) behind SLM is used for rotating the polarization. (**b**) Finally, the diffracted beam is focused into the sample, located under the focal plane of a 100× microscope objective (OBJ) and a chiral microstructure is achieved. (**c**) The beam modulated by designed holograms (first row) with different topological charges, which could be positive or negative, illuminates into the sample. Then the intensity pattern (second row) and optical micrograph of the polymer (third row) can be measured with a CCD. After developing, top-view (fourth row) and 45° tilted (fifth row) SEM images of chiral microstructures are realized. (Scale bar, 2 μm.)

A lower laser power is chosen to illuminate the sample for a particularly unambiguous fluorescent image. The induced fluorescent patterns caught by a CCD from the sample are consisting of a central bright dot and a 'doughnut' shape pattern, which are coincident with a plane wave and an optical vortex focused by an objective lens, respectively (Fig. 3c). A spiral pattern with $l$ ($\pm 3$, $\pm 4$ and $\pm 6$) lobes, whose direction is determined by the sign of topological charge, is evident in the polymer after illuminating with IVBs. More details of the chiral microstructure topography could be detected after developing. The distinct chiral microstructures with surface wrinkles have $l$ spiral lobes (top-view) and helices (45° tilted view) as observed in the SEM images.

Precise control over the shape and size of chiral microstructures needs to optimize the fabrication parameters. As shown in Figs. 4a-b, a series of chiral microstructures with three spiral lobes could be produced with fixed laser intensity and varying exposure time from 1 to 5 s. It is worth noting that although the diameters of chiral structures increase with a higher laser power or a longer exposure time, the shapes of chiral structures remain chirality and $l$ spiral lobes. An array of chiral microstructures with three spiral lobes in 2.4 mm × 2.4 mm square area is fabricated rapidly with irradiation power of 120 mW, and exposure time of 1 s (Figs. 4c-e). The total time is 4 hours, while the time may cost 20 days by conventional point-to-point scanning. When a 532 nm excitation light irradiates on the array, uniform spots pattern is diffracted on the received screen, indicating remarkable homogeneity of the large-area chiral microstructure array.

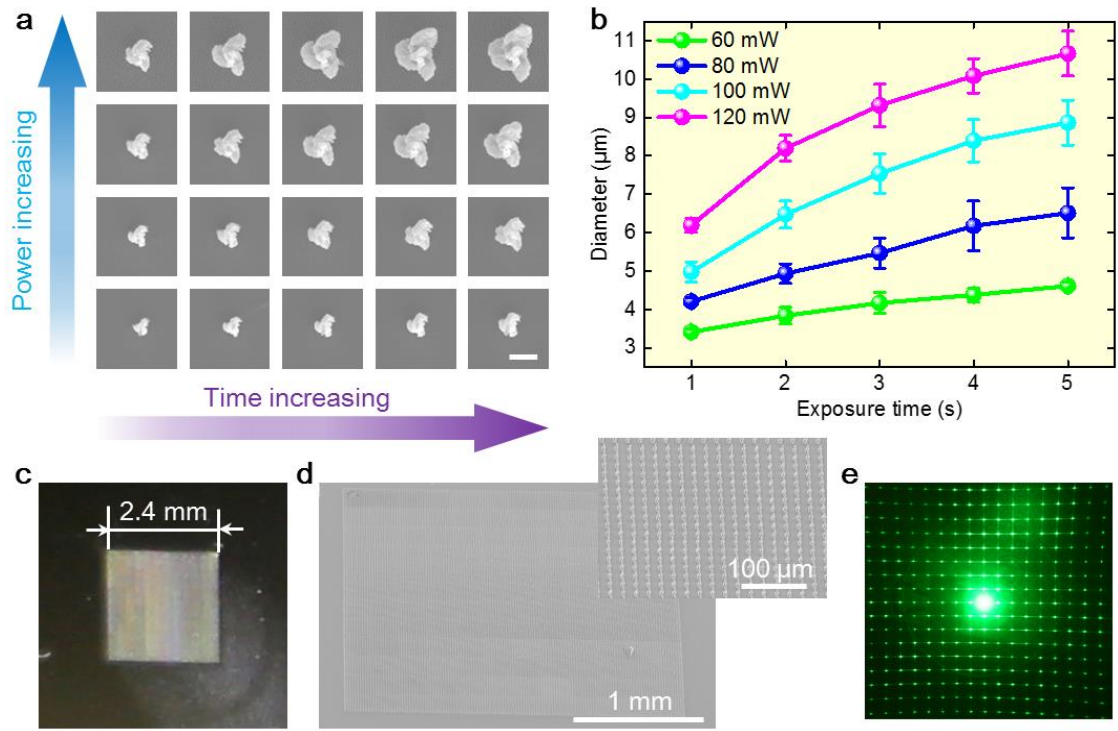

**Figure 4 | Large-area uniform chiral microstructures with controllable sizes by optimized parameters.** (**a**) SEM images of chiral microstructures with three spiral lobes achieved for different values of the laser power and exposure time. (Scale bar, 5 μm.) (**b**) Quantitative study on the diameters of chiral microstructure as a function of exposure time and power. (**c**) Optical image of a square area with 2D grating-like chiral microstructures has structural color, where the period is 20 μm. (**d**) SEM image of the arrayed chiral microstructures, inset is a magnified view. (**e**) Diffraction pattern fired by a laser with wavelength of 532 nm on the square area.

More complicated arrangements with three- four- and six-fold symmetric patterns, with 3, 4 and 6 spiral lobes, respectively, are achieved by controlling both holograms and positions[24] (Figs. 5a-f). It costs 4 second for fabricating a chiral microstructure with power 100 mW, and the total time for fabricating a 200 μm × 200 μm square is less than 8 mins. Microstructures arrays with inverse chirality can be readily produced on the same substrate by changing the holograms, such as letter 'L' with left handedness and letter 'R' with right handedness as illustrated in Figs. 5g, h. The desired patterns can be written on a flat surface or nonplanar microchannel[25] with precise location by a nano-positioning stage (Fig. 5i). Formation of a Fibonacci number pattern with left handedness microstructures can be grouped into thirteen counterclockwise spirals[26] (Fig. 5j). These images verify that IVB-based microfabrication is a robust and versatile approach for building 3D chiral microstructures, arraying in any desired 2D patterns.

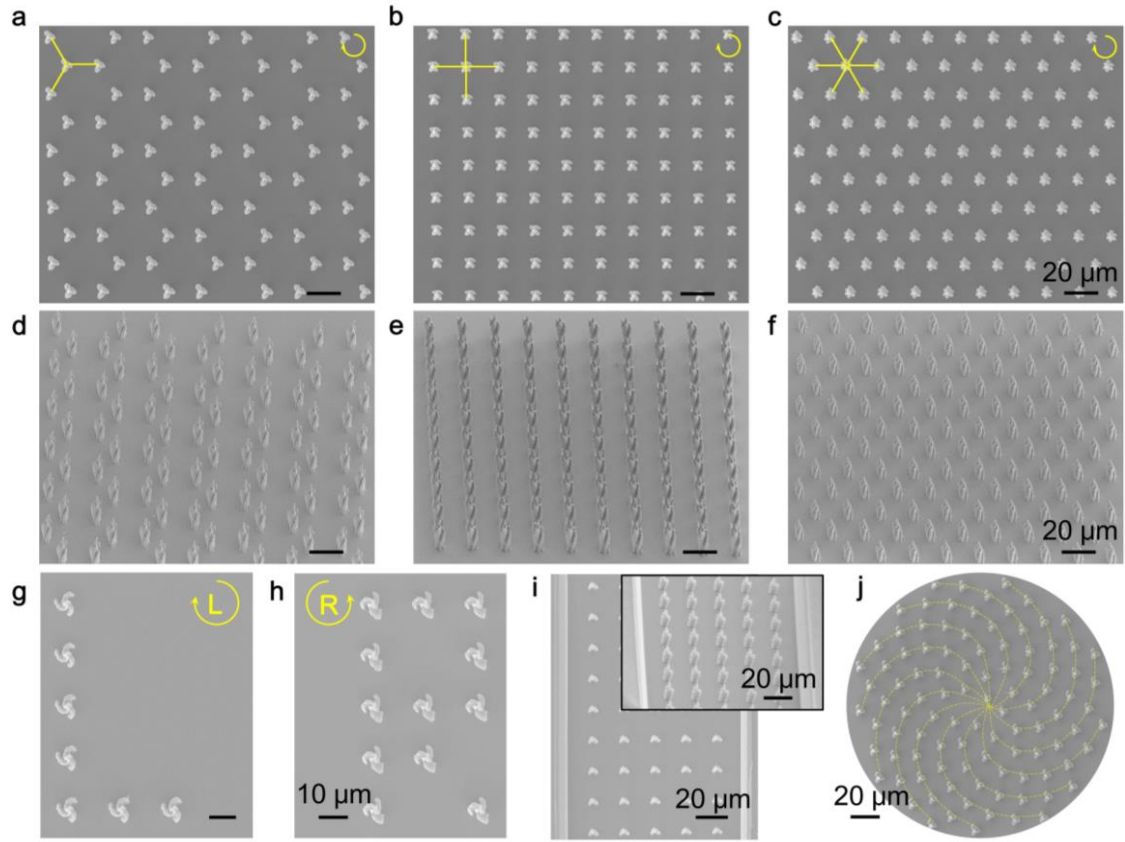

**Figure 5 | Diverse patterns prepared by chiral microstructures on flat and complex surface.** (**a-c**) top-view and (**d-f**) 45 °tilted SEM images of three-, four- and six-fold symmetric patterns with 3, 4 and 6 spiral lobes, respectively. (**g**, **h**) SEM images of microscale letters of 'L' (left) and 'R' (right) arranged with left- and right-handed microstructures, respectively. (**i**) Chiral microstructures integrated in a non-planar substrate (inset: 45 °tilted SEM image). (**j**) A Fibonacci number pattern is fabricated with topological charge $l = 3$.

**Orientation-controlled rotation of chiral microstructures by changing the phase value of plane wave.** As a common technique to visualize the OAM state of optical vortices, interference patterns could change the sense of rotation with the relative phase between the two interference beams. Specifically, objects can be trapped within the interference pattern of an optical vortex and a plane wave, consisting of $l$ spiral lobes, which is an available way to orientate multiple symmetric objects at the same time[21]. Since there is no free-space optics for creating a plane wave reference, the interference patterns are stable in our interfering system. Moreover, we can precisely rotate the chiral microstructure at any angle about its axis theoretically by changing the phase value of the plane wave, with topological charge $l = 3$ (see Fig. 6a). The phase value could be modified by changing the grey value in the CGHs, instead of mechanically tailoring the length of one arm of the interferometer. As dark gaps are generated by the IVBs, where azimuthal phase value equals to equiphase value on the IVHs, they will rotate by modifying the equiphase value. As a result, the inverse spiral lobes rotate visually on the optical field. Both the simulations (Fig. 6b) and measurements (Fig. 6c) of IVBs intensity profiles rotate the same angles with varying phase value of plane waves. Chiral microstructures fabricated by focused IVBs under the objective lens remain the rotating angle (see Fig. 6d). The equivalent phase area between azimuthal phase and equiphase in IVHs rotates by varying the phase of interfering plane wave from 0 to $2\pi$. The rotating angle of spiral lobes increases linearly with phase value of the plane wave by

$$\text{Ra} = \frac{360° \times \emptyset_0}{2\pi l} \quad (2)$$

where Ra is the rotating angle and $\emptyset_0$ is the equiphase value (Fig. 6e-f). An axisymmetric chiral microstructure fabricated with topological charge $l$ has spiral lobes with $l$-fold symmetries, causing a full rotation of $360°$ on the substrate.

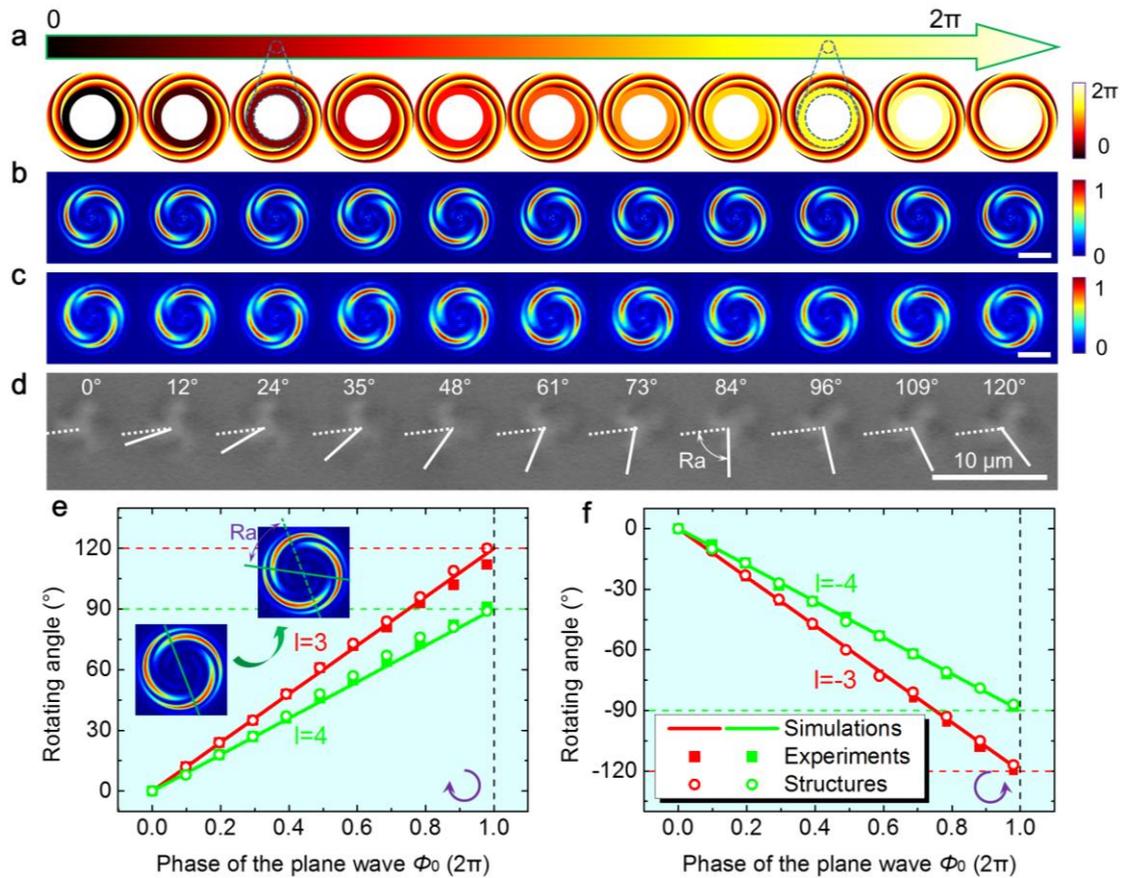

**Figure 6 | Orientation-controlled rotation of chiral microstructures by changing the phase value of plane wave.** (**a**) The phase of interfering plane wave increases from 0 to 2π indicated by an arrow with color gradient. The simulated intensity patterns (**b**) rotate in coincidence with the measurements (**c**). (Scale bar, 1 mm.) (**d**) Rotation of the chiral microstructures with $l = 3$. Rotating angle (Ra) of simulated and measured intensity patterns, and measured values of microstructures are as a function of the phase value of plane wave with $l$ = 3, 4 (**e**) and -3, -4 (**f**). The solid line represents the simulated rotating angle. The squares data and circles denote the measurements of optical intensity distributions and chiral microstructures, respectively. Vertical dashed lines indicate $\emptyset_0 = 2\pi$ for which the largest rotating angles are generated.

**Helical characterization and mechanism analysis of the 3D chiral microstructures.**
The chiral microstructures under illumination with IVBs feature $l$ spiral lobes on the transverse and helices on the optical axis. The $l$ spiral lobes could be associated with interference patterns, which have the same rotating angle in equation (2). There is not mass transport in this process, because the chirality is independent of the exposure time and applied power (see Fig.3). We cannot discuss the chirality by analyzing the transverse optical field only on the focal plane for 3D microstructures.

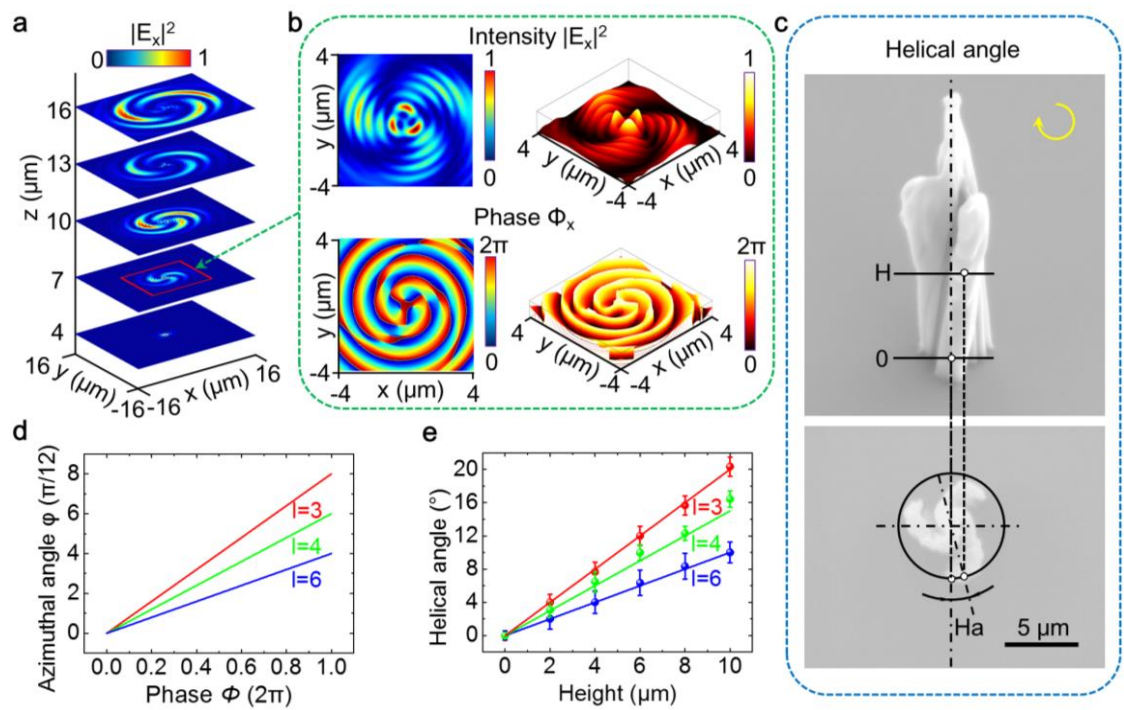

**Figure 7 | Helical characterization and mechanism analysis of the 3D chiral microstructures.**
(**a**) Simulated intensity distribution images of $E_x$ component in 3 μm steps at z-axis under the OBJ.

(**b**) The intensity and phase profiles are simulated on the plane, z = 7 μm away from the focal plane of OBJ. (**c**) The helical angle (Ha) of the chiral microstructure is defined as the rotating angle of cross-sections along the z-axis. (**d**) Azimuthal angle versus phase value in phase profile of optical vortices with $l$ = 3, 4 and 6, respectively. (**e**) Helical angle (Ha) are measured as a function of height (H), shown as points, and are compared to the values in Eq. (3), shown as solid lines. Error bars are the standard deviation of three measurements with the same sample.

Starting from this consideration, we have simulated the intensity profiles of IVB focused by the 100× objective lens on different distance from the focal plane as illustrated in Fig. 7a. The intensity and phase distributions of $E_x$ component at the distance z = 7 μm below the focal plane of objective lens are shown in Fig. 7b with both 2D and 3D formats. Three spiral lobes distribute in three-fold symmetry and the interference fringes are generated radially corresponding to the surface wrinkles of chiral microstructures. As no other device changes the azimuthal phase distribution in the beam propagation, the azimuthal optical phase gradients remain from optical vortices. The azimuthal phase distributions of optical vortices vary by $\emptyset = l\varphi$ (Fig. 7d). Three spiral lobes gradually shrink to the optical axis without obviously revolving in the transverse intensity profile far away from the focal plane. When arriving at the helical phase center, spiral lobes start to rotate predominately around the axis with a large azimuthal phase gradient (see Figs. S9, S10). For understanding the helices of chiral microstructures, Fig. 7c shows the definition of helical angle of the structure with a remarkable chirality. As forced by the azimuthal phase gradient, the spiral lobes rotate a helical angle (Ha) in the transverse estimated by

$$\text{Ha} = CH/l \qquad (3)$$

where $C$ is a constant characterizing the polymer and $H$ is the height of chiral microstructures (Fig. 7e). It means that the helices of chiral microstructures Ha/H ∝ $1/l$. Therefore the interference between optical vortices and plane waves generates intensity pattern consisting of $l$ spiral lobes, and the helical phase wavefronts keep them winded. Note that the coaxial interference with SLM in our experiments avoid rotating the spiral lobes of chiral microstructures undesirably by relative phase in an unstable environment. As a result, we demonstrate that the chiral microstructures with

$l$ spiral lobes record the information not only of the intensity patterns but also of the helical phase wavefronts.

## Discussion

We demonstrate 3D chiral microstructures with large-area and high precision fabricated in isotropic polymer by developing the coaxial interference of OAM beams and plane waves for the first time. The coaxial interference beams are realized by a simple and effective system for reflecting IVBs with integrated IVHs displayed on a SLM. 3D chiral microstructures with controllable spiral lobes and orientations are polymerized after illumination of IVBs. By designing the IVHs on SLM, chiral microstructures with controllable rotation angles could be achieved stably. We ascribe the spiral lobes to the interfering patterns and the chirality to the action of helical wavefronts. Our findings open a new promising way to fabricate chiral microstructures in an isotropic polymer with single pulse. Larger area array of chiral microstructures up to 1 cm$^2$ square are promisingly fabricated in five minutes with this technology by using a high power laser. Furthermore, uniform arrangement of chiral microstructures is potential to manufacture chiral metamaterials with applications in optoelectronic devices, analytical chemistry and circular dichroism spectroscopy[27]. As a method to generate flexibly stable interference patterns of optical vortices and plane waves, this technology also paves an innovative avenue in optical tweezers and optical communications.

## Methods

**Optical apparatus.** The femtosecond laser source is a mode-locked Ti:sapphire ultrafast oscillator(Coherent, Chameleon Vision-S) with central wavelength at 800 nm, pulse duration of 75 fs, and repetition rate at 80 MHz. The reflective liquid crystal SLM (Holoeye, Pluto NIR-2) has 1920 × 1080 pixels, with pixel pitch of 8μm and diagonal of 0.7 inch, on which CGHs with 256 grey levels responding to 2π phase value could display. Only the central portion of the SLM with 1080 × 1080 pixels is used for generating modulated beam and other pixels are assigned to zero as a reflective mirror[28]. The sample is mounted on a nano-positioning stage (Physik Instrument, E545) with a nanometer resolution and a 200 μm × 200 μm × 200 μm moving range to precisely locate microstructures.

**Sample preparation and characterization.** A commercially available zirconium-silicon hybrid sol-gel material (SZ2080) provided by IESL-FORTH (Greece) is used in our experiment, which is negligibly shrinkable during structuring compared with other photoresists. The pre-baking process used to evaporate the solvent in the SZ2080 is set to a thermal platform at 100 ℃ for half an hour. After polymerization under illuminated by femtosecond laser, the sample is developed in 1-propanol for 30 mins until all of the rest part without polymerization is washed away. The images are taken with a secondary electron scanning electron microscope (FEI Sirion200) operated at 10 keV, after the samples depositing ~10 nm gold.

**Numerical simulation at the plane A.** The electric field of incident light can be describe as

$$E^{in}(r,\varphi) = \begin{cases} A_0 \exp(-r^2/\omega^2)\exp[i\emptyset(r,\varphi)] & r_1 < r \leq r_3; \\ 0 & else. \end{cases} \quad (4)$$

where $A_0$ is the normalized constant. The focal length of lens L$_3$ is $f = 600$ mm and the distance between L$_3$ and the plane A is $d = 350$ mm. The phase transformation of lens L$_3$ could be written as

$$t(x_1, y_1) = \exp\left[-i\frac{k}{2f}(x_1^2 + y_1^2)\right] \quad (5)$$

Utilizing the Fresnel approximation of the scale diffraction theory, the diffracted field optical wave propagation from the plane L$_3(x_1, y_1)$ to the plane A$(x, y)$ could be described with the Fresnel diffraction integral[29]

$$U(x,y) = \frac{exp(ikz)}{i\lambda z} \iint_{-\infty}^{\infty} U^{in}(x_1, y_1) t(x_1, y_1) exp\left\{\frac{ik}{2z}[(x-x_1)^2 + (y-y_1)^2]\right\} dx_1 dy_1 \quad (6)$$

where $U(x, y)$ and $U^{in}(x_1, y_1)$ are the complex amplitude distribution of A$(x, y)$ and L$_3(x_1, y_1)$, respectively, $k = 2\pi/\lambda$ is the wavenumber, $\lambda$ is the wavelength, and $z = d$ is the propagation distance.

**Field distribution simulation under OBJ.** Our method starts by considering the IVBs with linear polarization focused by a high NA objective lens. The electromagnetic field near the focal spot could be derived by the vectorial diffraction theory[30-32] as

$$\vec{E}(r_2,\varphi_2,z_2) = \begin{bmatrix} E_x \cdot \hat{x} \\ E_y \cdot \hat{y} \\ E_z \cdot \hat{z} \end{bmatrix}$$

$$= iC[\hat{x}\ \hat{y}\ \hat{z}] \iint_\Omega \sin(\theta) E^{in}(\theta,\varphi) \sqrt{\cos\theta} \begin{bmatrix} 1+(\cos\theta-1)\cos^2\varphi \\ (\cos\theta-1)\cos\varphi\sin\varphi \\ \sin\theta\cos\varphi \end{bmatrix} \exp\{ik[z_2\cos\theta$$

$$+ r_2\sin\theta\cos(\varphi-\varphi_2)]\}d\theta d\varphi \qquad (7)$$

where $\vec{E}(r_2,\varphi_2,z_2)$ is the electric field vector at the point $(r_2,\varphi_2,z_2)$ being cylindrical coordinates at the focal plane, $C$ is a constant, $\theta$ is the convergence, $\theta_{max}$ is determined by the NA of the objective and $k$ is the wave number of the incidence.

## Acknowledgement:

This work is supported by National Science Foundation of China (Nos. 51275502, 61475149, 51405464, 91223203, 91423103 and 11204250), Anhui Provincial Natural Science Foundation (No. 1408085ME104), National Basic Research Program of China (No. 2011CB302100), the Fundamental Research Funds for the Central Universities (WK2480000002), and "Chinese Thousand Young Talents Program".